\documentclass[12pt]{iopart}

\usepackage{graphicx}
\begin{document}

\title[]{Structure and stability of the magnetic solar tachocline}

\author{G R\"udiger$^1$ and L L Kitchatinov$^{1,2}$ }

\address{$^1$~Astrophysikalisches Institut Potsdam, An der Sternwarte 16,
              D-14482, Potsdam, Germany}
\address{$^2$~Institute for Solar-Terrestrial Physics, PO Box
             4026, Irkutsk 664033, Russia }
\ead{gruediger@aip.de; kit@iszf.irk.ru}
\begin{abstract}
Rather weak fossil  magnetic fields in the  radiative core can produce the solar tachocline if the  field is poloidal and almost horizontal in the tachocline region, i.e. if the field is confined within the core. This particular   field geometry is shown to result from a shallow
({\lower.4ex\hbox{$\;\buildrel <\over{\scriptstyle\sim}\;$}}1\,Mm) penetration  of the meridional flow existing in the  convection zone into the radiative core. I.e., two conditions are  crucial for  a magnetic tachocline theory: (i) the presence of meridional flow of a few meters per second at the base of the convection zone, and (ii) a magnetic diffusivity inside the tachocline smaller than $10^8$~cm$^2$s$^{-1}$. Numerical solutions for both  confined poloidal fields and the resulting tachocline structures  are presented. We find that the tachocline thickness runs as $B_\mathrm{p}^{-1/2}$ with the poloidal field amplitude  falling below 5\% of the solar radius for $B_\mathrm{p} > 5$ mG. The resulting toroidal field amplitude inside the tachocline of about 100~G  does not depend on the $B_\mathrm{p}$. The hydromagnetic stability of the  tachocline is only briefly discussed. For the hydrodynamic stability of latitudinal differential rotation we found   that the  critical 29\%  of the 2D theory of Watson (1981) are reduced to only 21\% in 3D for marginal modes of about 6 Mm radial scale.
\end{abstract}


\section{Introduction}
\begin{figure}
   \centering
   \includegraphics[width=9cm,height=7cm]{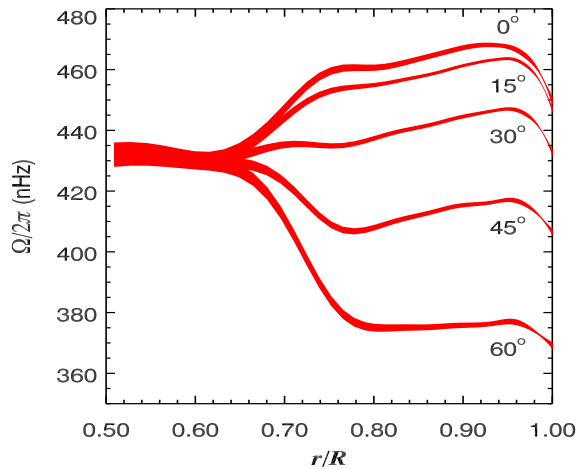}
   \caption{The internal solar rotation from helioseismology.
            The lines are marked by correspondent latitude values
	    (courtesy NSF's National Solar Observatory).
              }
   \label{f1}
\end{figure}

The tachocline is a thin shell inside the Sun where the  rotation pattern changes strongly. Beneath the tachocline, the solar rotation is rather uniform. Above the tachocline, the rotation rate varies with latitude to decrease from equator to pole by about 30\%. The transition from differential to rigid rotation detected by helioseismology (Wilson \etal 1997; Kosovichev \etal 1997; Schou \etal 1998) is shown in figure~\ref{f1}.  The tachocline thickness is about  5\% of the solar radius, its midpoint radius is  (0.692 $\pm$ 0.005)$R_\odot$, and it is slightly prolate in shape (Kosovichev 1996; Antia \etal 1998; Charbonneau \etal 1999b). The tachocline is located  mostly if not totally beneath the base of convection zone at $R_\mathrm{in} = 0.713R_\odot$ (Christensen-Dalsgaard \etal 1991; Basu \& Antia 1997).

The given picture of the internal solar rotation suggests  that some coupling must  exist between the base of convection zone and the inner core. Otherwise, the rotational braking of the Sun on evolutionary time-scales would produce  a rapidly rotating core (Dicke 1970). More than that, an even  stronger link between low and high latitudes   operates immediately below the base of convection zone in order to produce the rotation profile within the tachocline.
The magnetic theory which explains the tachocline structure as a consequence  of a weak internal fossil  magnetic field simultaneously provides both the links (R\"udiger \& Kitchatinov 1997; MacGregor \& Charbonneau 1999). The present  paper summarizes the basic  ideas of the magnetic theory and describes new results of numerical MHD to  model  the tachocline and probing the internal field structure. Other approaches has been recently reviewed by Gilman (2005) and Garaud (2007).

 Section~\ref{model} formulates the  equations of the tachocline model and shows their solution for prescribed geometry of the fossil  magnetic field within  the radiative  core. The role of meridional flow penetrating the uppermost region of the radiative core from the  convection zone is discussed in section~\ref{confinement}. In this section, the eigenmodes of internal poloidal field are computed with inclusion of  the penetrating flow. Tachocline models with consistently defined internal field modes are discussed in section~\ref{tacho}. In the final section~\ref{Watson} a first step is presented to understand the stability problem of the solar tachocline. In a {\em hydrodynamical} approach the 2D Watson theory for the stability of latitudinal differential rotation is reformulated in 3D with surprising results.  The  modes  with small radial wavelengts are most unstable so that it is indeed essential for  tachocline theories to take into account all the radial stratifications.
\section{An heuristic approach}\label{estimations}
\begin{figure}
   \centering
   \includegraphics[width=6cm,height=5cm]{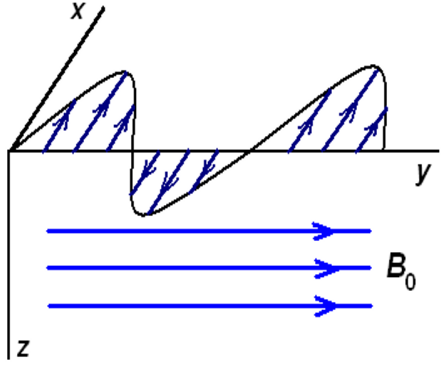}
   \caption{A shear flow is prescribed at $z=0$. Its penetration into
   	the region of $z>0$ where the  magnetic field $\bi{B}_0$ exists 
	mimics the  magnetic tachocline formation.
              }
   \label{f2}
\end{figure}

The basic parameters of the tachocline theory can be estimated as follows.
The estimates  are made with a simplified shear flow model in Cartesian geometry sketched in figure~\ref{f2}. The plane of $z=0$ mimics the bottom of the convection zone where a shear flow $\bi{u} = \left(U(y,z),0,0\right)$ imitating the differential rotation is prescribed as 
\begin{equation}
     U = U_0\sin\left(ky\right)\ \ \mathrm{at}\ z=0.
     \label{1}
\end{equation}
The Cartesian coordinates $x, y$ and $z$ correspond to azimuth, latitude and depth beneath the convection zone. An  uniform (poloidal) field $\bi{B}_0$ along the $y$-axis is prescribed in the radiative zone. The shear flow penetrates the $z>0$ region and produces an $x$-component $B(y,z)$ of the field so that 
\begin{equation}
     \bi{B} = \left(B(y,z), B_0, 0\right) .
     \label{2}
\end{equation}
can be written.
The  equations for the stationary (toroidal) $x$-components of both magnetic and velocity fields then read
\begin{equation}
    \nu\left(\frac{\partial^2 U}{\partial y^2} +
    \frac{\partial^2 U}{\partial z^2}\right) +
    \frac{B_0}{\mu_0\rho}\frac{\partial B}{\partial y} = 0, \qquad
    \eta\left(\frac{\partial^2B}{\partial y^2} +
    \frac{\partial^2 B}{\partial z^2}\right) +
    B_0\frac{\partial U}{\partial y} = 0 ,
    \label{3}
\end{equation}
where $\nu$ is the viscosity and $\eta$ the  magnetic diffusivity. The boundary condition for the 
flow is \eref{1}. For the toroidal field we impose the vacuum condition $B = 0$ at $z=0$  motivated by the very large turbulent magnetic diffusivity inside the convection zone compared with microscopic diffusivity of the radiative interior. The remaining two conditions require $U$ and $B$ to vanish for $z\rightarrow\infty$.

The solution of the equations \eref{3} simply reads
\begin{eqnarray}
    U(y,z) &=& U_0\exp\left(-\lambda_1 k z\right)
    \cos\left(\lambda_2 k z\right)\sin ky ,
    \nonumber \\
    B(y,z) &=& \sqrt{\mu_0\rho}\ U_0 \mathrm{Pm}^{1/2}
    \exp\left(-\lambda_1 k z\right)
    \sin\left(\lambda_2 k z\right)\cos ky ,
    \label{4}
\end{eqnarray}
where $\mathrm{Pm} = \nu/\eta$ is the magnetic Prandtl number and the parameters
\begin{eqnarray}
   \lambda_1 &=& \left(1 + \mathrm{Ha}^2\right)^{1/4}
   \cos\left(\case{1}{2}\mathrm{arctanHa}\right),
   \nonumber \\
   \lambda_2 &=& \left(1 + \mathrm{Ha}^2\right)^{1/4}
   \sin\left(\case{1}{2}\mathrm{arctanHa}\right)
   \label{5}
\end{eqnarray}
depend on the Hartmann number
\begin{equation}
    \mathrm{Ha} = \frac{B_0}{k\sqrt{\mu_0\rho\nu\eta}} ,
    \label{6}
\end{equation}
which is the basic  parameter of the whole theory.

With Ha\,=\,0, the equation \eref{5} gives $\lambda_1 = 1$, $\lambda_2 = 0$, and \eref{4} converts into the nonmagnetic solution, $B = 0,\ U = U_0\rme^{-kz}\sin ky$. The resulting shear flow penetrates deep inside the $z >0$ region. No slender tachocline can be formed without  magnetic fields.
But with solar parameters one finds $\mathrm{Ha} \simeq 10^7 B_0$ ($B_0$  in G) just below the convection zone. Very large Ha can thus be expected. For this case, $\lambda_1 = \lambda_2 = \sqrt{\mathrm{Ha}/2}$, and the solution \eref{4} provides two important consequences:

(i) The tachocline thickness, $D_\mathrm{tach}$, is strongly reduced compared to its nonmagnetic value, $D_0$,
\begin{equation}
     D_\mathrm{tach} = D_0\sqrt{\frac{2}{\mathrm{Ha}}} \simeq
     \sqrt{\frac{D_0}{B_0}}\left(\mu_0\rho\nu\eta\right)^{1/4} .
     \label{7}
\end{equation}
The tachocline is so thin because its extension in $z$-direction is only due to viscous stress while the smoothing in $y$-direction is performed by the much stronger Maxwell stress.

(ii) The amplitude of the toroidal magnetic field in the steady tachocline does not depend on the poloidal field strength, i.e. 
\begin{equation}
   B = U_0 \sqrt{\mu_0\rho\ \mathrm{Pm}} .
   \label{8}
\end{equation}
The ratio of magnetic to kinetic energy in the tachocline equals Pm. In terms of the Alfv\'en velocity, $V_\mathrm{A} = B/\sqrt{\mu_0\rho}$, one finds
\begin{equation}
   V_\mathrm{A} = \sqrt{\mathrm{Pm}}\ U_0 
   \label{9}
\end{equation}
 predicting $B \sim 1000$~G for the Sun.

After \eref{7} even a weak  poloidal field of only  $B_0 \sim 10^{-3}$~G can reduce $D_\mathrm{tach}$ below 5\% of the solar radius.  Equation \eref{7} does also apply when $B_0$ is not an internal field of the radiative core.  In case the  poloidal field of the solar cycle  diffuses into the core (Forg\'acs-Dajka \& Petrovay 2002)  $\nu$ and $\eta$ must be replaced by the turbulent diffusivity values. As their magnetic Prandtl number is not much smaller than  unity  the resulting toroidal magnetic field after \eref{9} becomes {\em much} stronger than 1000 G.
\section{Tachocline model in spherical geometry}\label{model}
The tachocline equations will be formulated for  axial symmetry. Then the magnetic field can  be expressed in terms of $A$ and $B$ in accord to 
\begin{equation}
    \bi{B} = {\mathbf e}_\phi B +
    \nabla\times\left({\mathbf e}_\phi\frac{A}{r\sin\theta}\right) ,
    \label{10}
\end{equation}
where $r,\theta$ and $\phi$ are spherical coordinates and $\mathbf e_\phi$ is the azimuthal unit vector. In the next section  we show that meridional flow is only significant for the structure of the poloidal field. It can be neglected in the tachocline equations below. Here the only mean flow is the differential   rotational with $\Omega=\Omega(r,\theta)$.

If the  poloidal field is given, the tachocline is described by the equations for the toroidal field $B$ and the angular velocity $\Omega$
\begin{eqnarray}
   &&\frac{\eta}{r}
   \frac{\partial}{\partial\theta}\left(\frac{1}{\sin\theta}
   \frac{\partial\left( B\sin\theta\right)}{\partial\theta}\right)
   + \frac{\partial}{\partial r}\left(\eta
   \frac{\partial\left( Br\right)}{\partial r}\right) =
   \frac{\partial\Omega}{\partial\theta}
   \frac{\partial A}{\partial r} -
   \frac{\partial\Omega}{\partial r}
   \frac{\partial A}{\partial\theta} ,
   \nonumber \\
   &&\frac{\rho\nu}{\sin^3\theta}\frac{\partial}{\partial\theta}
   \left( \sin^3\theta\frac{\partial\Omega}{\partial\theta}\right) +
   \frac{1}{r^2}\frac{\partial}{\partial r}\left(
   r^4\rho\nu\frac{\partial\Omega}{\partial r}\right) =
   \nonumber \\
   && \qquad = \frac{1}{\mu_0 r^2\sin^3\theta}\left(r
   \frac{\partial A}{\partial r}
   \frac{\partial\left( B\sin\theta\right)}{\partial\theta} -
   \sin\theta\frac{\partial A}{\partial\theta}
   \frac{\partial\left( Br\right)}{\partial r}\right).
   \label{11}
\end{eqnarray}
The rotation profile  at the top boundary is produced by the  convection zone for which  the expression
\begin{equation}
     \Omega = 2.9\left(1 - 0.15\cos^2\theta\right)\
     \mu\mathrm{rad\ s}^{-1}\quad \mathrm{at}\ \ r = R_\mathrm{in} ,
     \label{12}
\end{equation}
is used derived by Charbonneau \etal (1999a).
The remaining  boundary conditions are the vacuum condition for toroidal field on the top and the conditions for the regularity of the fields for $r=0$:
\begin{equation}
     B|_{r=R_\mathrm{in}} = B|_{r=0} =
     \left.\frac{\partial\Omega}{\partial\theta}\right|_{r=0} = 0 .
     \label{13}
\end{equation}

\begin{figure}
   \centering
   \includegraphics[width=8.8cm,height=6cm]{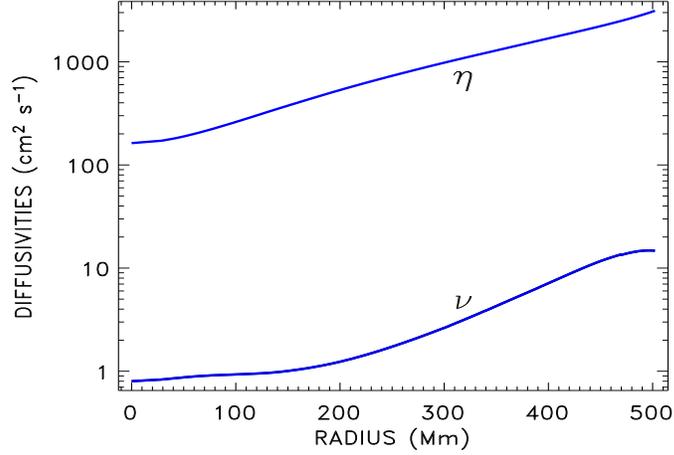}
   \caption{Radial profiles of magnetic diffusivity \eref{14}
            and viscosity \eref{15} calculated with   the solar structure
	    model by Stix \& Skaley (1990).
              }
   \label{f3}
\end{figure}

The used diffusivity profiles are shown in \fref{f3}. We do not expect turbulence in the radiative core and apply, therefore,  the  microscopic magnetic diffusivity
\begin{equation}
    \eta = 10^{13} T^{-3/2}\ \mathrm{cm}^2\mathrm{s}^{-1} ,
    \label{14}
\end{equation}
and the viscosity
\begin{eqnarray}
   &&\nu = \nu_\mathrm{micro} + \nu_\mathrm{rad},
   \nonumber \\[0.1cm]
   &&\nu_\mathrm{micro} = 1.2\cdot 10^{-16} \frac{T^{5/2}}{\rho}\
   \mathrm{cm}^2\mathrm{s}^{-1},\quad
   \nu_\mathrm{rad} = 2.5\cdot 10^{-25}\frac{T^4}{\kappa\rho}\
   \mathrm{cm}^2\mathrm{s}^{-1}
   \label{15}
\end{eqnarray}
({\it cf.} Kippenhahn \& Weigert 1994) including   molecular
($\nu_\mathrm{micro}$) and radiative ($\nu_\mathrm{rad}$) parts;  $\kappa$ is opacity. 

The equations \eref{11} with their boundary conditions \eref{12}, \eref{13}  provide tachocline solutions if the poloidal field is known. For a very first view the  field model
\begin{equation}
    A = B_\mathrm{p}\frac{r^2}{2}\left(1-\frac{r}{R_\mathrm{in}}
    \right)^q\sin^2\theta,\quad q>1
    \label{16}
\end{equation}
is used  to probe  whether  slender tachoclines can indeed be found. $B_\mathrm{p}$ is the free amplitude of the poloidal field.
\begin{figure}[h]
   \centering{
   \includegraphics[width=5cm,height=5cm]{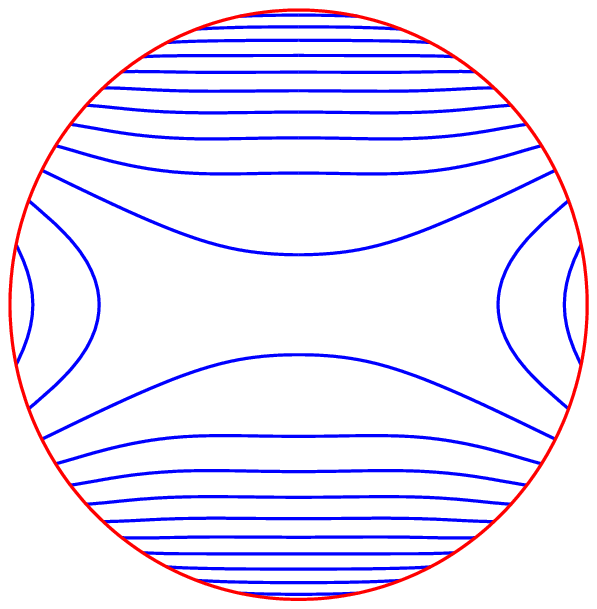}
   \hspace{0.3cm}
   \includegraphics[width=5cm,height=5cm]{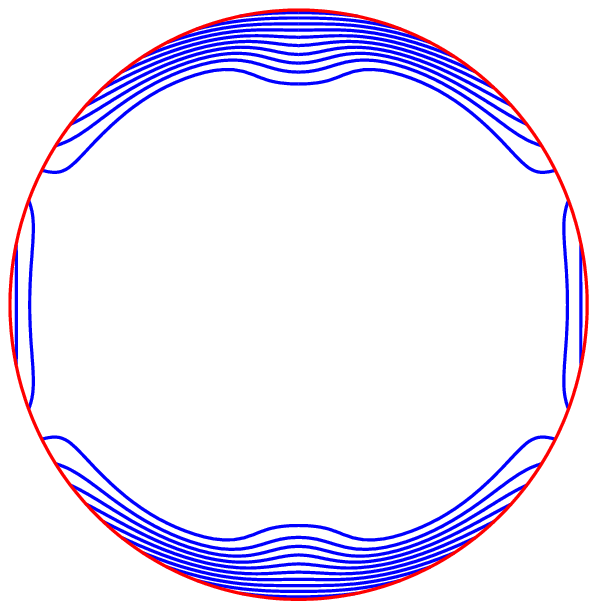}
   }
   \caption{Isolines of the  angular velocity of rotation inside the radiative core
            for the poloidal field model of \eref{16} with $q=5$.
	    The outer circle is the base of the convection zone. {\it Left} :  $B_\mathrm{p} = 0$. No  tachocline exists. {\it Right}:  $B_\mathrm{p} = 0.01$~G. Tachocline exists.
              }
   \label{f4}
\end{figure}
\Fref{f4} compares the  rotation pattern in a radiative zone computed with either zero magnetic field and with $B_\mathrm{P} = 0.01$~G. $B_\mathrm{P}$ is the \emph{maximum} strength of poloidal field inside the core. The maximum is attained in the core center. The field in the tachocline region is much weaker. Even this  rather  weak field suffices to produce a very strict  and slender tachocline.
\section{Magnetic field confinement by meridional flow}\label{confinement}
Our tachocline model is not  yet consistent. It works with a prescribed poloidal field. We have shown that the  model is not very sensitive to the poloidal field amplitude. Even a weak field can produce tachoclinic structures. The model is, however, very sensitive to the field geometry. The tachocline can be found only if the field lines are almost horizontal near the top of radiative zone. Steady rotation and magnetic fields can be realized in highly conducting fluids only with constant angular velocity along the field lines (Ferraro 1937). Therefore, the open field geometry with  field lines crossing the boundary between convection and radiative zones is not appropriate for  tachocline formation. The poloidal field prescribed by  \eref{16} is of a closed type. The field can, however, change to an open structure due to magnetic diffusion (Brun \& Zahn 2006). We shall show  in the following that already a small {\em  penetration} of meridional flow from the convection zone into the radiative core produces the confined field geometry required for the tachocline formation 
(see Kitchatinov \& R\"udiger 2006). If the electric conductivity in the core is high enough compared to that of the  convection zone then the flow has massive consequences and the poloidal field lines in the transition zone become parallel to the meridional flow ({\it cf.} Mestel 1999).
\subsection{Penetration of meridional circulation}
\begin{figure}
   \centering{
   \hspace{2.0cm}
   \includegraphics[width=6cm,height=5.7cm]{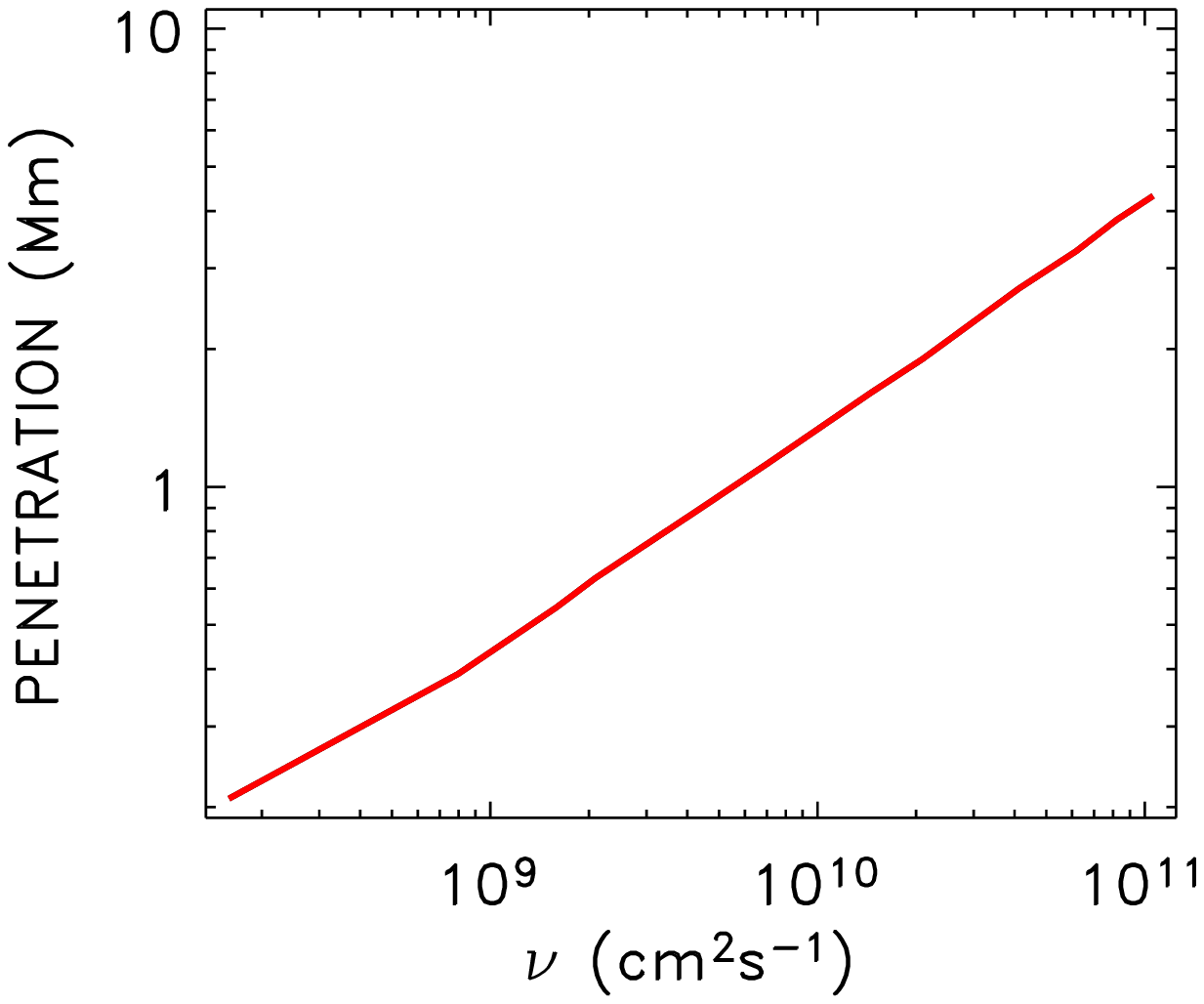}
   \hspace{0.3cm}
   \includegraphics[width=6cm]{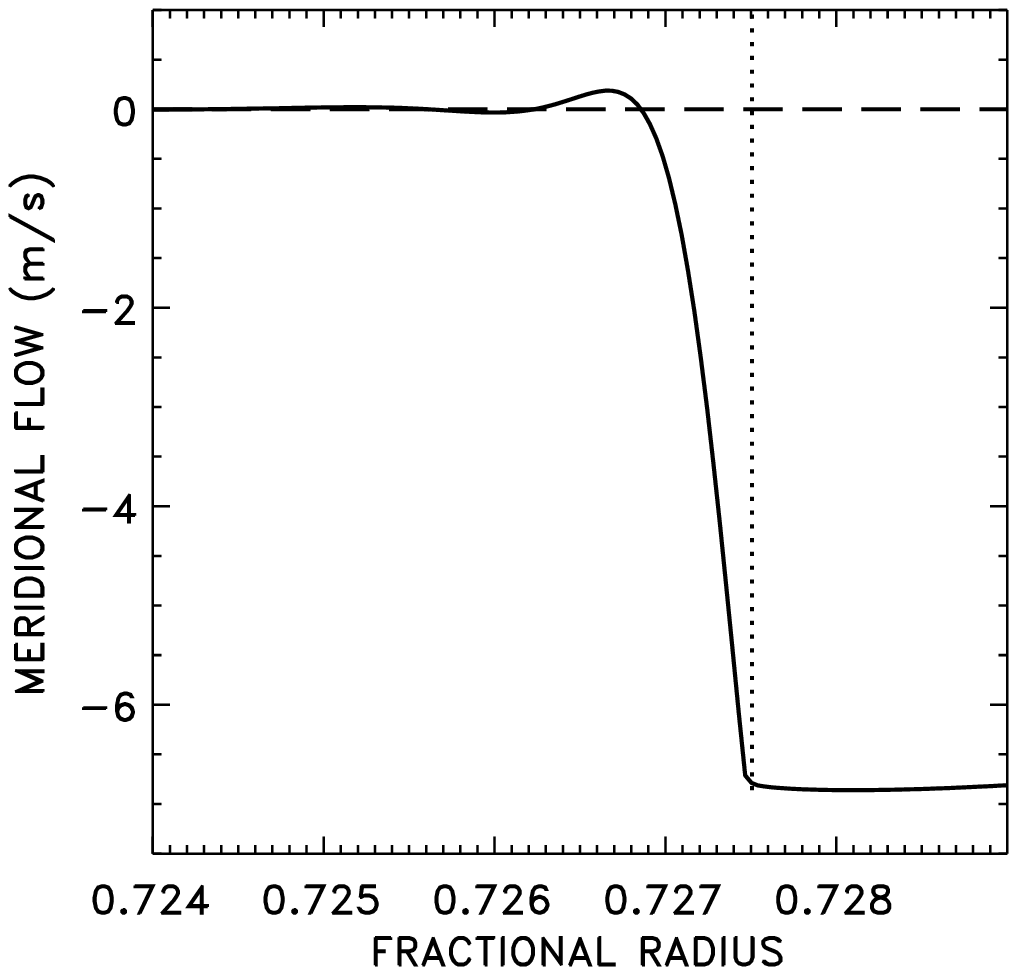}
   }
   \caption{{\it Left:} Penetration depth as function of  viscosity of the radiative core.
   	{\it Right:} Meridional velocity in the penetration region
	at $45^\circ$ latitude for
	$\nu = 1.1\cdot 10^9$~cm$^2$s$^{-1}$. The vertical dotted line
	marks the bottom of the convection zone.
              }
   \label{f5}
\end{figure}

A global poleward flow is observed at the solar surface (Komm \etal 1993)  persisting to a depth of at least 12~Mm (Zhao \& Kosovichev 2004). There must be a return flow towards the equator somewhere deeper. Theoretical models  indeed predict an equatorward flow of $\sim 10$\,m\,s$^{-1}$ at the base of convection zone (Kitchatinov \& R\"udiger 1999; Miesch \etal 2000; Rempel 2005). This flow can penetrate beneath the bottom of the convection zone  into the radiative core (\fref{f5}). This  penetration has been discussed recently in relation to dynamo models for solar activity (Nandy \& Choudhuri 2002; Gilman \& Miesch 2004; R\"udiger \etal 2005).

The penetration results from the viscous drag imposed by the meridional flow at the base of convection zone on the fluid beneath, and this is opposed by Coriolis force. The thickness of such an  Ekman layer results as $D_\mathrm{pen} \sim \sqrt{\nu/\Omega}$ (see Gilman \& Miesch 2004). The plot in the left of \fref{f5} is indeed approximated by
$D_\mathrm{pen} \simeq 2.3\cdot \sqrt{\nu/\Omega}$ or, equivalently, by
\begin{equation}
    D_\mathrm{pen} = 1.4\cdot \sqrt{\nu}\cdot 10^3\ \mathrm{cm}
    \label{17}
\end{equation}
for solar parameters\footnote{$D_\mathrm{pen}$ of \fref{f5} was defined as distance from the convection zone bottom  to the position where the meridional flow falls to zero}.
The penetration distance is about 100\,m  for microscopic viscosity  and remains shorter than 1\,Mm for any reasonable value of eddy viscosity. The distance is small even compared to the tachocline depth, i.e. $D_\mathrm{pen} \ll D_\mathrm{tach}$.

This shallow penetration is, nevertheless, highly  important for the geometry of the internal poloidal field. The ratio of the time $\tau_\mathrm{diff} = D^2_\mathrm{pen}/\eta$ for the magnetic field diffusion across the penetration layer to the characteristic time of induction  of a meridional magnetic field from a radial one, $\tau_\mathrm{s} = D_\mathrm{pen}/u^\mathrm{m}$, gives the magnetic Reynolds number
\begin{equation}
   \mathrm{Rm} = \frac{\tau_\mathrm{diff}}{\tau_\mathrm{s}} =
   \frac{D_\mathrm{pen}u^\mathrm{m}}{\eta} \simeq
   \frac{\mathrm Pm}{\sqrt{\nu}}\cdot 10^6
   \label{18}
\end{equation}
(with $u^\mathrm{m}\simeq 10$~m\,s$^{-1}$  for the meridional velocity).
This Reynolds number is large, Rm$\sim 10^3$, for  microscopic diffusivities but  it remains above this limit  with eddy diffusivities up to $10^6$~cm$^2$s$^{-1}$.
The large ratio of  \eref{18} means that the latitudinal  field inside the penetration layer is large compared to the radial field component. With other words,  the field has just the confined geometry required for the tachocline formation. Only with eddy diffusivities of $10^{12}$~cm$^2$s$^{-1}$ the Reynolds number \eref{18} sinks to unity, so that only in this case the influence of penetration on the internal field geometry is too  weak.

The ratio of $\tau_\mathrm{diff}$ to the advection time $\tau_\mathrm{adv} = R_\mathrm{in}/u^\mathrm{m}$ 
\begin{equation}
    \frac{\tau_\mathrm{diff}}{\tau_\mathrm{adv}} =
    \frac{D_\mathrm{pen}}{R_\mathrm{in}}\ \mathrm{Rm}  \simeq
    0.03\ \mathrm{Pm} \ll 1 ,
    \label{19}
\end{equation}
is small independently of whether microscopic or eddy diffusivity is used. If  the tachocline is stable in the hydrodynamic regime (see section 6) the microscopic diffusivities should be used. The small ratio \eref{19} justifies the neglect  of the  meridional flow in the tachocline equations \eref{11}. The diffusion time, $\tau_\mathrm{diff}$, is too short as could  the penetration layer be dynamo-relevant. The configuration of  the internal poloidal field is probably the only process for which the penetration of the meridional circulation into the stable radiative zone  is important. 
\subsection{A model of the poloidal field}
For both axisymmetric flows  and fields   the poloidal magnetic field equation decouples from the tachocline equations \eref{11} and  reads
\begin{equation}
    \frac{\partial A}{\partial t} =
   -\frac{u_\theta}{r}\frac{\partial A}{\partial\theta}
   - u_r \frac{\partial A}{\partial r}
   + \eta\frac{\partial^2 A}{\partial r^2}
   + \frac{\eta}{r^2}\sin\theta\frac{\partial}{\partial\theta}
   \left(\frac{1}{\sin\theta}\frac{\partial A}{\partial\theta}\right) ,
   \label{20}
\end{equation}
where $\bi{u}$ is the velocity field. The vacuum boundary condition for the poloidal field can be applied on the top of the radiative zone  because of the very large turbulent diffusivity in the convection zone,
\begin{equation}
   \frac{\partial A}{\partial r} =
   \left(\frac{\partial A}{\partial r}\right)_\mathrm{vac}\quad
   {\rm at}\ \ \ r = R_\mathrm{in}.
   \label{21}
\end{equation}
The condition is usually formulated in terms of a  Legendre polynomials expansion
\begin{eqnarray}
   A\left( r,\theta\right) &=& \sin\theta\sum\limits_{n=1}^\infty
   A_n\left( r\right) P_n^1\left(\cos\theta\right) ,
   \nonumber \\
   \left(\frac{\partial A}{\partial r}\right)_\mathrm{vac} &=&
   - \frac{\sin\theta}{r} \sum\limits_{n=1}^\infty
   n A_n\left( r\right) P_n^1\left(\cos\theta\right) .
   \label{22}
\end{eqnarray}
The other condition is $A=0$ at $r=0$.

The velocity $\bi{u}$ can be written with its stream function $\psi$ as
\begin{equation}
  \bi{u} = \left(\frac{1}{\rho r^2\sin\theta}
  \frac{\partial\psi}{\partial\theta},
  -\frac{1}{\rho r\sin\theta}\frac{\partial\psi}{\partial r},
  r\sin\theta\ \Omega\right).
    \label{23}
\end{equation}
The meridional flow in the bulk of the radiative core is very slow. The characteristic time of the Eddington-Sweet circulation even exceeds the solar age (Tassoul 2000). Only the flow penetrating from convection zone is significant for the tachocline. The stream function, $\psi(r,\theta)$, of the penetrating flow drops to zero at a small depth of order $D_\mathrm{pen}$ inside the core. The function is, however, finite at the top boundary of the core where it scales as
\begin{equation}
    \psi\left(R_\mathrm{in},\theta\right) =
    u^\mathrm{m}\rho R_\mathrm{in}D_\mathrm{pen}\hat\psi\left(\theta\right) ,
    \label{24}
\end{equation}
where $\hat\psi$ is dimensionless function of order unity. $u^\mathrm{m}$ is the meridional velocity amplitude at the boundary.

The penetration depth is so small even compared to tachocline thickness that we are motivated to integrate equation \eref{20} across the penetration layer instead of resolving this  layer explicitely. Such an  integration reformulates the top boundary condition which now reads
\begin{equation}
   R_\mathrm{in}\frac{\partial A}{\partial r} -
   \mathrm{Rm}
   \frac{\hat\psi\left(\theta\right)}{\sin\theta}
   \frac{\partial A}{\partial\theta} =
   R_\mathrm{in}\left(\frac{\partial A}{\partial r}\right)_\mathrm{vac},
   \label{25}
\end{equation}
where Rm is the magnetic Reynolds number \eref{18} and the RHS is defined in \eref{22}. For  $\rm Rm=0$ \eref{25} reduces to the
vacuum condition \eref{21}. The true Reynolds number is 
$\mathrm{Rm}\sim 10^3$.

Penetrating meridional flow is now included via the condition \eref{25} and poloidal field structure can be found by solving eigenvalue problem for diffusion equation,
\begin{equation}
   -\frac{A}{\tau} =
   \eta\frac{\partial^2 A}{\partial r^2}
   + \frac{\eta}{r^2}\sin\theta\frac{\partial}{\partial\theta}
   \left(\frac{1}{\sin\theta}\frac{\partial A}{\partial\theta}\right) ,
   \label{26}
\end{equation}
where eigenvalue $\tau$ is the time of resistive decay of normal modes.
The results discussed below were obtained in computations with the simplest stream function,
\begin{equation}
   \hat\psi \left(\theta\right) =
   -\sin\theta\bar{P}^1_2\left(\cos\theta\right),
   \label{27}
\end{equation}
where $\bar{P}^1_2$ is the normalized Legendre polynomial.

The penetrating flow is expected to confine the internal poloidal field within the radiative core. The parameter
\begin{equation}
 \delta\phi = \frac
 {\mathrm{max} | A\left( r,\theta\right)|_{r = R_\mathrm{b}}}
 {\mathrm{max} | A\left( r,\theta\right)|_{r \leq R_\mathrm{b}}},
 \label{28}
\end{equation}
is used to estimate  the field confinement.  The $\delta\phi$-parameter \eref{28} estimates the ratio of the magnetic flux through the surface of the core to the characteristic value of the flux within the core.

The linear equations \eref{26} and \eref{25} define the field structure but not its amplitude. The amplitude  $B_\mathrm{p}$ of the poloidal field is a free  parameter of the model.
\subsection{Normal modes of the internal field}\label{eigenmodes}
\begin{figure}[h]
   \centering{
   \hspace{2.5cm}
   \includegraphics[width=4cm,height=4cm]{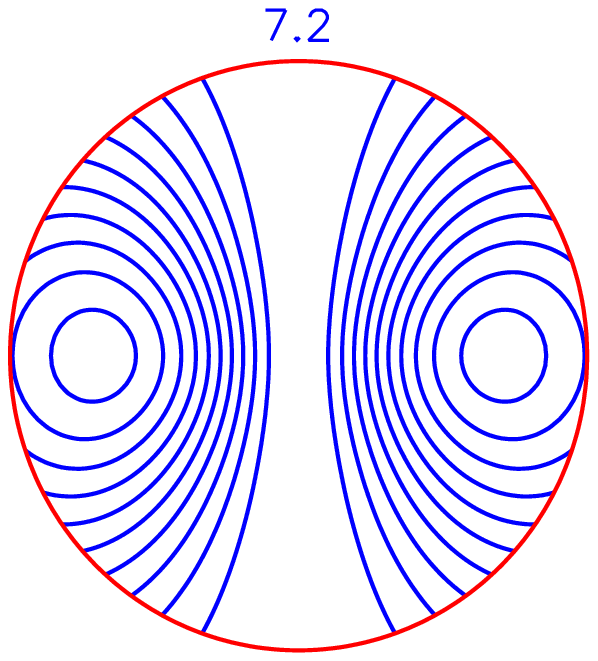}
   \hspace{0.2cm}
   \includegraphics[width=4cm,height=4cm]{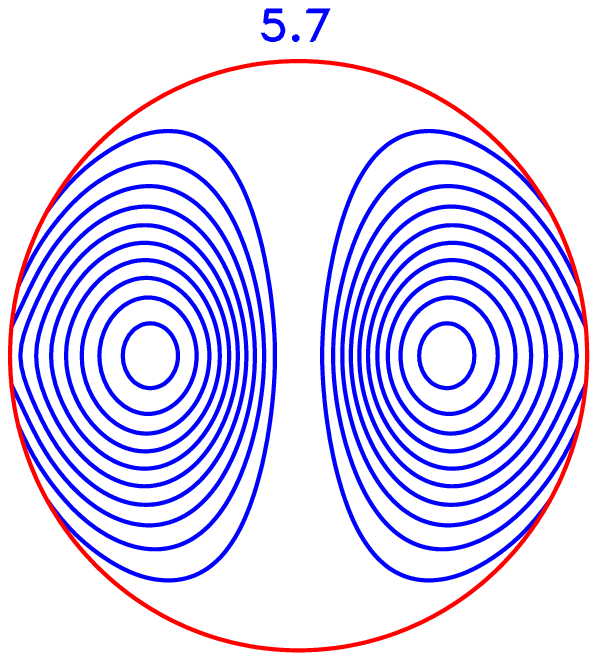}
   \hspace{0.2cm}
   \includegraphics[width=4cm,height=4cm]{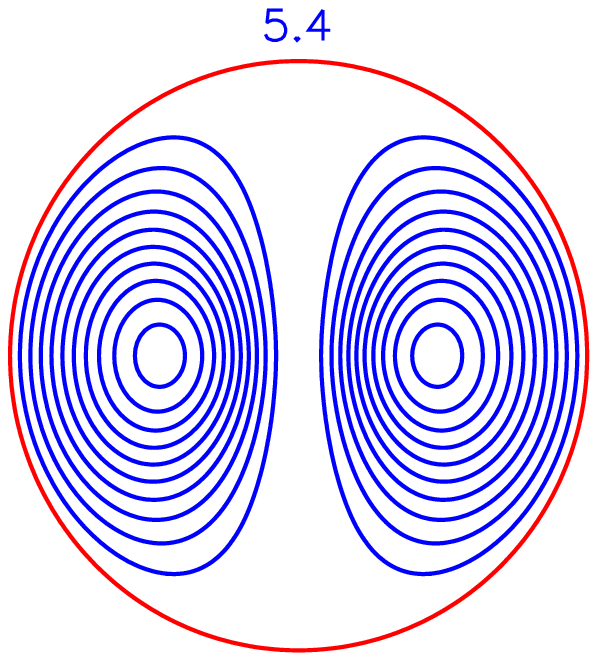}
   }
   \caption{Field lines for the longest-living dipolar modes of poloidal
   	field for different Rm. The Reynolds number
   	varies as $\rm Rm=0,\ 10,\ 1000$ from left to right.
	The outer circle is the base of convection zone. The decay times in Gyr are marked at the tops.
              }
   \label{f6}
\end{figure}

\begin{figure}[h]
   \centering
   \includegraphics[width=9cm, height=5.5cm]{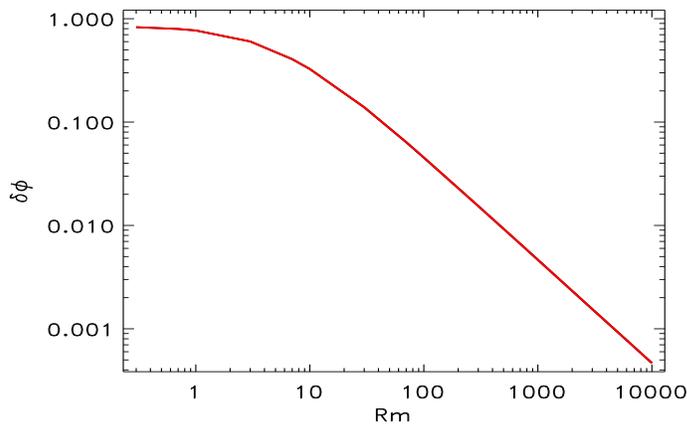}
   \caption{Confinement parameter \eref{28} for the longest-living
   	dipolar modes as function of the magnetic
	Reynolds number Rm.
              }
   \label{f7}
\end{figure}

\begin{figure}
   \centering{
   \hspace{2.5cm}
   \includegraphics[width=4cm]{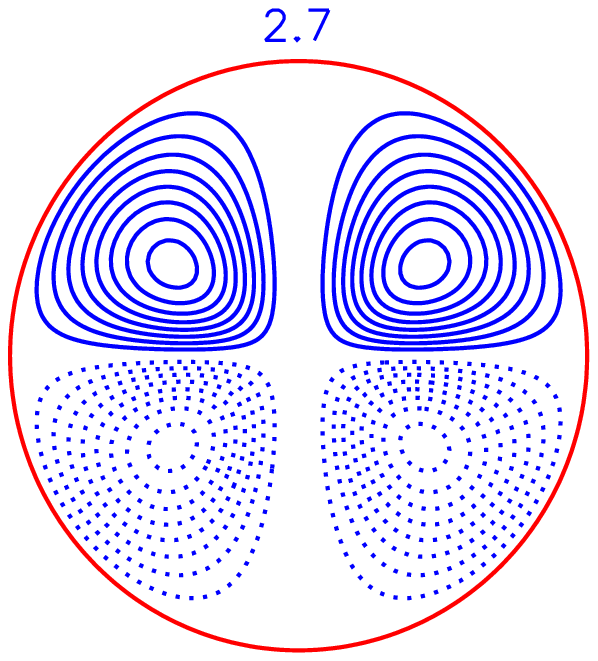}
   \hspace{0.2cm}
   \includegraphics[width=4cm]{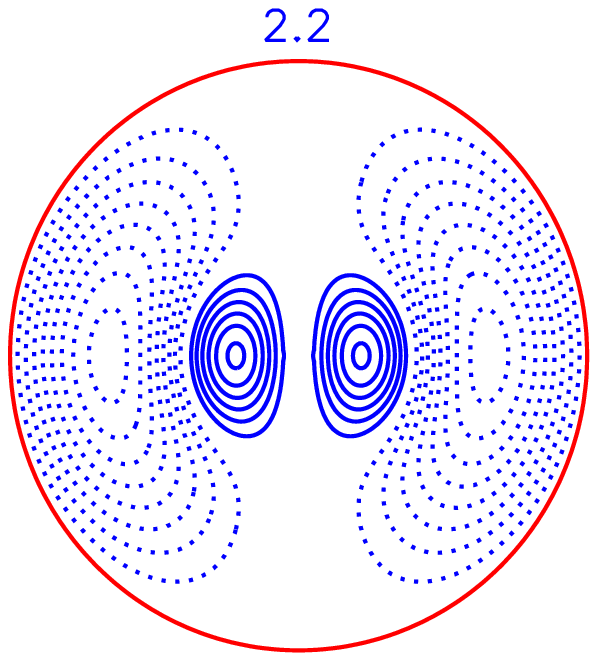}
   \hspace{0.2cm}
   \includegraphics[width=4cm]{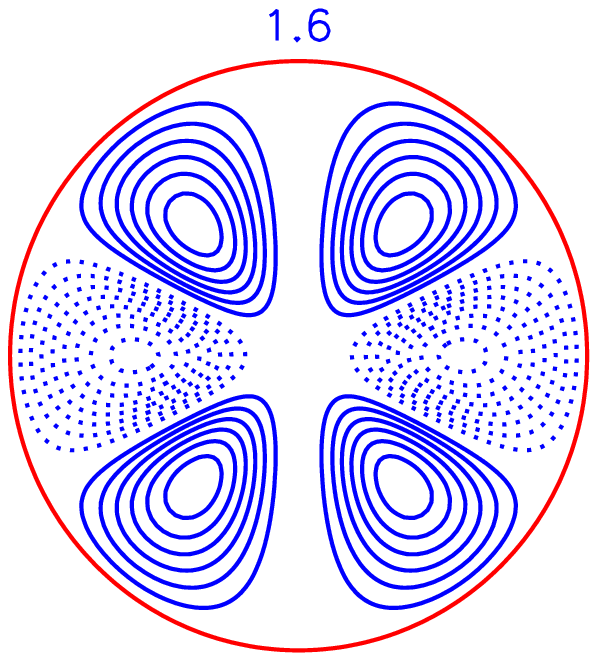}
   }
   \caption{Normal modes of internal field following the dipolar mode
            of \fref{f6}  for shorter  lifetimes.
	    The decay times (in Gyr) are shown on the top.
	    The confinement parameter \eref{28} varies as
	    0.0051, 0.0073, 0.0026 from left to right. $\rm Rm=1000$.
              }
   \label{f8}
\end{figure}

 How  the structure of the internal field changes towards a confined geometry as Rm increases is shown in  \fref{f6}. The internal field computed with $\rm Rm=0$ has an open structure. Even a moderate flow with $\rm Rm=10$ changes the field considerably towards the confined geometry. The internal field  for the solar value $\rm Rm=1000$   shown in \fref{f6}  is almost totally confined.  Less than 1\% of magnetic flux belongs to \lq open' field lines in this case  (\fref{f7}).
The figures \ref{f6} and \ref{f7} only display  the most slowly decaying dipolar modes. Other normal modes have shorter scales  in radius or latitude and also shorter decay times. The three modes following the longest-living dipole  ordered for  decreasing lifetimes are shown in \fref{f8}. The higher-order modes also have a confined structure. The decay times of the given normal modes are long enough to allow tachocline computations with steady poloidal fields.

For sufficiently high Rm the  direction of the penetrating meridional flow does not play a role. The flow from equator to poles also makes  efficient confinements of the internal field. The effect can be understood as a   magnetic field expulsion from the region of circulating motion (Weiss 1966). The transition from convection zone to radiative core was treated as a sharp boundary in our model. A smooth decrease of eddy diffusivity in convection zone towards its base  may  also contribute to the field confinement via the diamagnetic effect of inhomogeneous turbulence ({\it cf.} Krause \& R\"adler 1980).
\section{Tachocline models}\label{tacho}
\begin{figure}
   \centering{
   \hspace{2.5cm}
   \includegraphics[width=4cm]{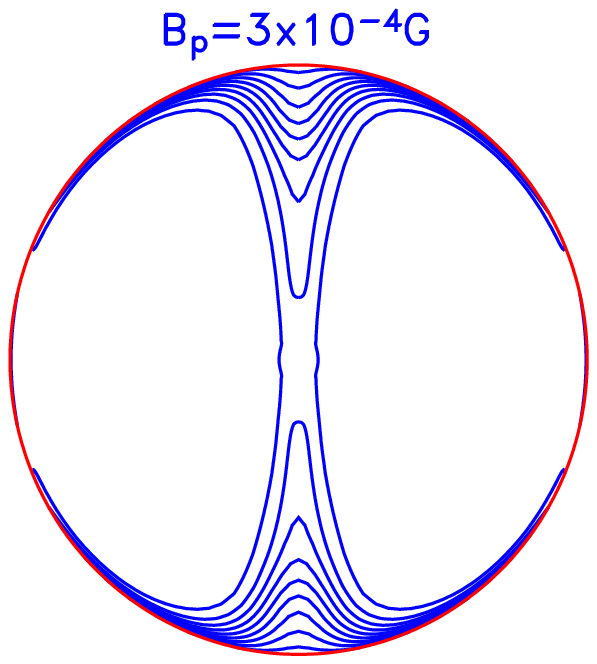}
   \hspace{0.2cm}
   \includegraphics[width=4cm]{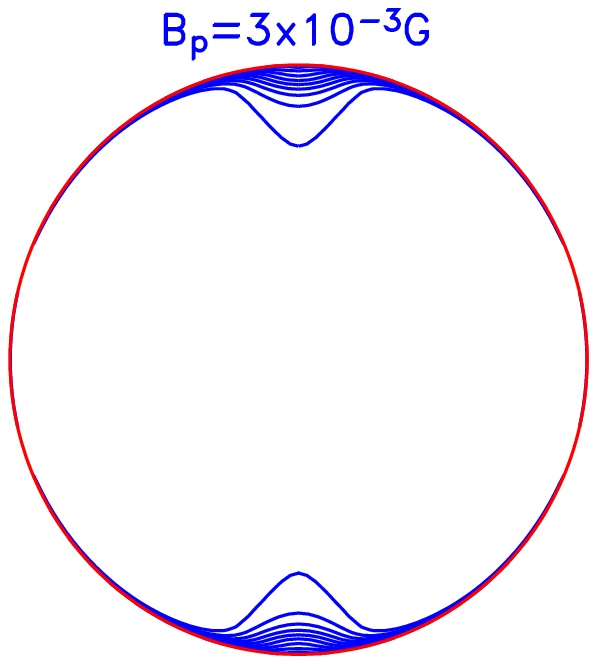}
   \hspace{0.2cm}
   \includegraphics[width=4cm]{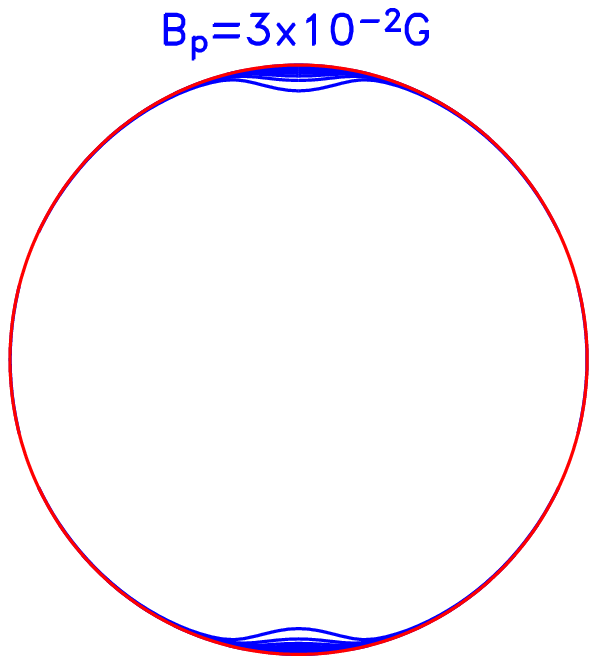}
   }
   \caption{Angular velocity isolines in the radiative core
            computed with dipolar internal fields for
	    $\rm{Rm} =1000$ for various  poloidal field
	    amplitudes (marked at  the top).
              }
   \label{f9}
\end{figure}

\begin{figure}
   \centering
   \includegraphics[width=9cm, height=5.5cm]{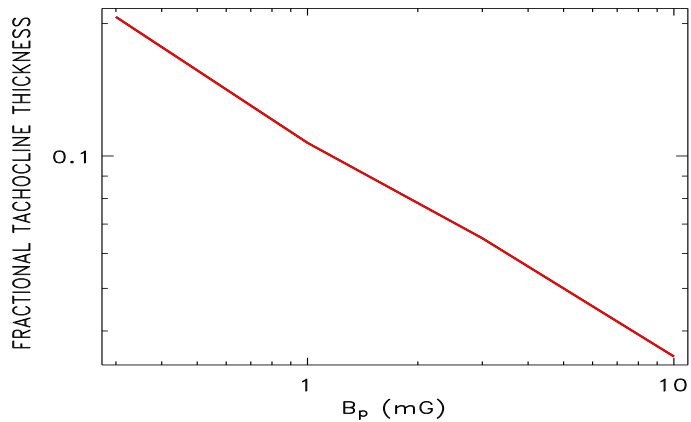}
   \caption{The fractional tachocline thickness 
            as function of the amplitude,
            $B_\mathrm{p}$, of the internal poloidal field.
            Rm$=1000$.
              }
   \label{f10}
\end{figure}

\Fref{f9} shows the angular velocity distributions inside the core  computed with dipolar eigenmodes of internal field of various strengths.
Dependence of the tachocline thickness on poloidal field amplitude is shown in \fref{f10} ($D_\mathrm{tach}$ was defined as depth of exponential decrease of equator-to-pole difference in angular velocity). Even a very weak field, $B_\mathrm{p} \sim 10^{-3}$~G, can produce tachocline. The depth $D_\mathrm{tach}$ drops below 5\% of the solar radius for $B_\mathrm{p} > 5\ \mathrm{mG}$.

The tachocline thickness in relation to the amplitude to the prescribed fossil field is given in  \fref{f10}. It is very   close to the rough estimate \eref{7}. The simulations also confirm the above finding   that the  toroidal field amplitude, $B_\mathrm{t} = 100...200$~G, hardly varies while the  poloidal field amplitude changes by several orders of magnitude.

Also the higher order modes given in  \fref{f8}  also produce tachoclinic structures  provided that the fields have the confined structure which is always the case for Rm$=100$ or larger.

Figures \ref{f9} and \ref{f10} are valid for  $\rm Rm=1000$. Already $\rm Rm=100$ suffices for tachocline formation, but for smaller Rm the  poloidal field remains  \lq too open' for the formation of a  tachocline.  With other words,  a meridional flow with  minimum 1~m\,s$^{-1}$ amplitude at the base of convection zone is necessary for  the success of  the magnetic tachocline theory. The rather shallow penetration of the meridional flow of the convection zone into the radiative core influences the internal field geometry strongly enough and in such a way that the field becomes appropriate for the tachocline formation. We suggest that a  layer thiner than 1~Mm beneath the convection zone where a meridional flow of 1-10~m\,s$^{-1}$ enters the radiative core is responsible for the existence  of the solar tachocline.
\section{Hydrodynamic stability: the Watson approach in 3D}\label{Watson}
Latitudinal differential rotation can be unstable even without magnetic field if the shear  $\partial\Omega/\partial \theta$ is positive and sufficiently strong (Watson 1981). The critical value of 29\% latitudinal shear found by Watson for the nonaxisymmetric mode with $m=1$ resulted from a theory with strong stratification but without any radial velocity and radial wavelengths.   The value has  also appeared   in a 3D numerical examination of marginal stability of a shell rotating fast enough with the rotation law 
\begin{equation}
    \Omega = \Omega_0\left( 1 - a\cos^2\theta\right), 
    \label{W28}
\end{equation}
but of fully incompressible material (Arlt \etal 2007).   The critical shear increases to  higher values, however,  if the real  rotation law (including its radial variations) of the solar tachocline is adopted. In this case the superrotation observed in the equatorial region of the rotation law strongly stabilizes the shear instability.

\begin{figure}
   \centering
   \includegraphics[width=8.8cm, height=6.0cm]{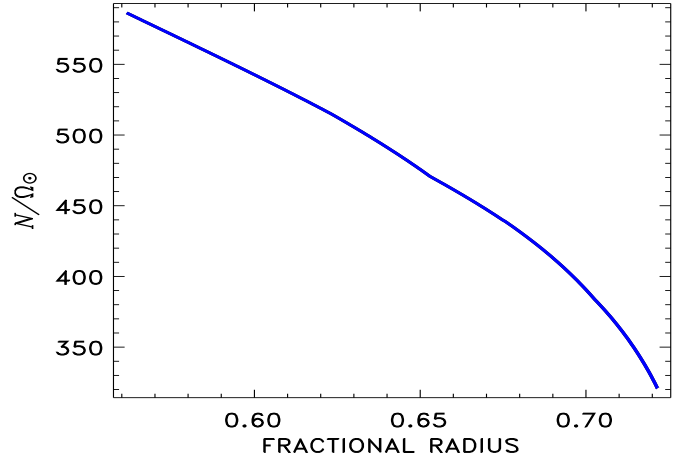}
   \caption{The buoyancy frequency \eref{W2}  in the upper radiative core 
            after the solar model of Stix \& Skaley (1990).
              }
   \label{Wf1}
\end{figure}

In this section  the Watson approach with strong density stratification (for $m=1$) is extended to 3D, i.e.  to the inclusion of  finite radial velocities and finite radial wavelengths.
The stabilizing effect of a subadiabatic stratification on  the mean flow  is characterized by the buoyancy frequency,
\begin{equation}
   N^2 = \frac{g}{C_\mathrm{p}}\frac{\partial S}{\partial r} ,
   \label{W2}
\end{equation}
where $S = C_\mathrm{v}\ln \left( P/\rho^\gamma\right)$ is the specific entropy of ideal gas. The frequency is very large  in the solar radiative core (see figure~\ref{Wf1}).
The larger the  $N$  the more  the  radial fluctuations  are suppressed by the buoyancy force. Radial velocities should therefore be small. 
Our stability analysis is local in the radial dimension, i.e. we use Fourier modes $\exp{\mathrm{i}kr}$ with $kr \gg 1$. The equation system remains, however, global in horizontal dimensions.

Instabilities of rotating shear flows  are fast. Their growth rates are of order $\Omega$ or smaller. In the short-wave approximation the velocity field can be assumed as divergence-free, $\mathrm{div}{\bi u}' = 0$.  The  next assumption concerns the  pressure. Local thermal disturbances occur at constant pressure so that $\rho'/\rho = -T'/T$ or $S' = -C_\mathrm{p}\rho'/\rho$ where perturbations  have been  marked by dashes. This assumption is again justified by the incompressible  nature of the fluid.
We start from the linearized equations  of velocity, i.e.
\begin{eqnarray}
 \frac{\partial{\bi u}'}{\partial t}+
 \left({\bi u}\cdot\nabla\right){\bi u}'
 + \left({\bi u}'\cdot\nabla\right){\bi u}
 =-\left(\frac{1}{\rho}\nabla p\right)' + \nu\Delta{\bi u}'
  \label{W5}
\end{eqnarray}
and entropy
\begin{equation}
  \frac{\partial S'}{\partial t} + {\bi u}\cdot \nabla S' +
  {\bi u}'\cdot\nabla S= \frac{C_\mathrm{p}\chi}{T}\Delta T' .
  \label{W7}
\end{equation}
The basic flow   is the rotation law \eref{W28}.

 Perturbations of velocity and pressure are now expressed in terms of scalar potentials like
\begin{equation}
  {\bi u}' = \frac{{\bi e}_r}{r^2}\hat{L}P
  - \frac{{\bi e}_\theta}{r}\left(\frac{1}{\sin\theta}
  \frac{\partial T}{\partial\phi} +
  \frac{\partial^2 P}{\partial r\partial\theta}\right)
  +\frac{{\bi e}_\phi}{r}\left(\frac{\partial
  T}{\partial\theta} -
  \frac{1}{\sin\theta}\frac{\partial^2 P}
  {\partial r\partial\phi}\right) ,
 \label{W3}
 \end{equation}
 with the operator
 \begin{equation}
  \hat{L} = \frac{1}{\sin\theta}\frac{\partial}{\partial\theta}
  \sin\theta\frac{\partial}{\partial\theta} +
  \frac{1}{\sin^2\theta}\frac{\partial^2}{\partial\phi^2}.
  \label{W3a}
\end{equation}
The identities
\begin{eqnarray}
   r \left({\bi r}\cdot\nabla\times{\bi u}'\right) &=&
   \hat{L} T,\ \ \
   r^3\left({\bi r}\cdot\nabla\times\nabla\times{\bi u}'\right) =
   \left(\hat{L} + r^2\frac{\partial^2}{\partial r}\right)
   \hat{L}P,
   \label{W8}
\end{eqnarray}
are used to reformulate the relations in terms of the potentials.

The perturbations are considered as Fourier modes for  time, azimuth and radius in the form  $\mathrm{exp}\left(\mathrm{i}(-\omega t + m\phi + kr)\right)$. For an instability  the eigenvalue $\omega$ must  possess a positive imaginary part. Only the highest order terms in $kr$ for the same variable have been considered.
The pressure term in \eref{W5} can be formulated  as 
\begin{eqnarray}
   {\bi r}\cdot\nabla\times\nabla\times\left(\frac{1}{\rho}
   \nabla p\right)' &=&
   -{\bi r}\cdot\nabla\times\left(\frac{1}{\rho^2}
   (\nabla\rho)\times(\nabla p)\right)' =
   \nonumber \\
   \frac{\bi r}{C_\mathrm{p}}\cdot\nabla\times\left(\frac{1}{\rho}
   (\nabla S)\times(\nabla p)\right)' &=&
   -\frac{\bi r}{C_\mathrm{p}}\cdot\nabla\times
   \left({\bi g}\times\nabla S'\right) =  \frac{g}{rC_\mathrm{p}}
   \hat{L}S' .
   \label{W9}
\end{eqnarray}

In order to denormalize the  variables  time is measured in units of $\Omega_0^{-1}$ and    velocities are scaled with  $r\Omega_0$. The  remaining  variables are
\begin{equation}
   V = \frac{k}{\Omega_0 r^2}\ P,\ \
   W = \frac{1}{\Omega_0 r^2}\ T,\ \
   s = \frac{\mathrm{i} k r g}{C_\mathrm{p} r N^2}\ S' ,\ \ \ \
   \hat\Omega = \frac{\Omega}{\Omega_0}.
   \label{W10}
\end{equation}
Then the  equation for the poloidal flow reads
\begin{eqnarray}
   \hat\omega\left(\hat{L}V\right) &=&
   -\hat{\lambda}^2\left(\hat{L}s\right)
   - \mathrm{i}\frac{\epsilon_\nu}{\hat{\lambda}^2}\left(\hat{L}V\right)-
   \nonumber \\
   &-& 2\mu\hat\Omega\left(\hat{L}W\right)
   - 2\left(1-\mu^2\right)\frac{\partial\left(\mu\hat\Omega\right)}
   {\partial\mu}\frac{\partial W}{\partial\mu}
   -2 m^2\frac{\partial\hat\Omega}{\partial\mu} W+
   \nonumber \\
   &+& m\hat\Omega\left(\hat{L}V\right)
   + 2m\frac{\partial\left(\mu\hat\Omega\right)}{\partial\mu} V
   + 2m\left(1-\mu^2\right)\frac{\partial\hat\Omega}{\partial\mu}
   \frac{\partial V}{\partial\mu},
   \label{W11}
\end{eqnarray}
with $\mu = \cos\theta$ and the normalized radial wavelength
\begin{equation}
   \hat\lambda = \frac{N}{\Omega_0 k r}.
   \label{W12}
\end{equation}
The first term on the RHS describes the stabilizing effect of the  stratification. It  vanishes for small $\hat\lambda$. Apart from this  stabilizing buoyancy term, the wavelength only exists in diffusive terms. The second term of the RHS  includes the action of   viscosity. Here we have used the quantities 
\begin{equation}
  \epsilon_\nu = \frac{\nu N^2}{\Omega_0^3 r^2},\ \ \ \ \ \ \ \ \ \ \ \ \ 
  \epsilon_\chi = \frac{\chi N^2}{\Omega_0^3 r^2}
  \label{W13}
\end{equation}
for the diffusive parameters $\nu$ and $\chi$. Now the limit $N \to 0$ can no longer be used for the transition to nonstratified  but viscous media. The numerical values for the solar tachocline are 
\begin{equation}
\epsilon_\nu = 2\cdot 10^{-10},\ \ \ \ \ \ \ \ \ \ \ \ \ 
  \epsilon_\chi = 10^{-4}.
\label{num}
\end{equation} 
The second and the following lines in \eref{W11} describe the influences of the basic rotation. Note that only latitudinal derivatives of $\Omega$ appear. 

The complete system of  equations also includes the equation for toroidal flow,
\begin{eqnarray}
  \hat\omega\left(\hat{L}W\right) =
  - \mathrm{i}\frac{\epsilon_\nu}{\hat\lambda^2}\left(\hat{L}W\right)
  + m\hat\Omega\left(\hat{L}W\right)
  - mW\frac{\partial^2}{\partial\mu^2}
  \left(\left(1-\mu^2\right)\hat\Omega\right)+
  \nonumber \\
  +\left(\hat{L}V\right)\frac{\partial}{\partial\mu}
  \left(\left(1-\mu^2\right)\hat\Omega\right)
  +\left(\frac{\partial}{\partial\mu}\left(\left(1-\mu^2\right)^2
  \frac{\partial\hat\Omega}{\partial\mu}\right) -
  2\left(1-\mu^2\right)\hat\Omega\right)\frac{\partial V}{\partial\mu},
  \label{W14}
\end{eqnarray}
and the entropy equation
\begin{equation}
   \hat\omega s = - \mathrm{i}\frac{\epsilon_\chi}{\hat\lambda^2} s
   + m\hat\Omega s + \hat{L}V .
   \label{W17}
\end{equation}
  Note that for the Reynolds number ${\rm Re} = \Omega r^2/\nu$ follows
\begin{equation}
{\rm Re} = \frac{N^2/\Omega^2}{\epsilon_\nu},
\label{Re}
\end{equation}
which  with \eref{num}  is  very large.

\begin{figure}
   \centering
   \includegraphics[width=10.0cm, height=7cm]{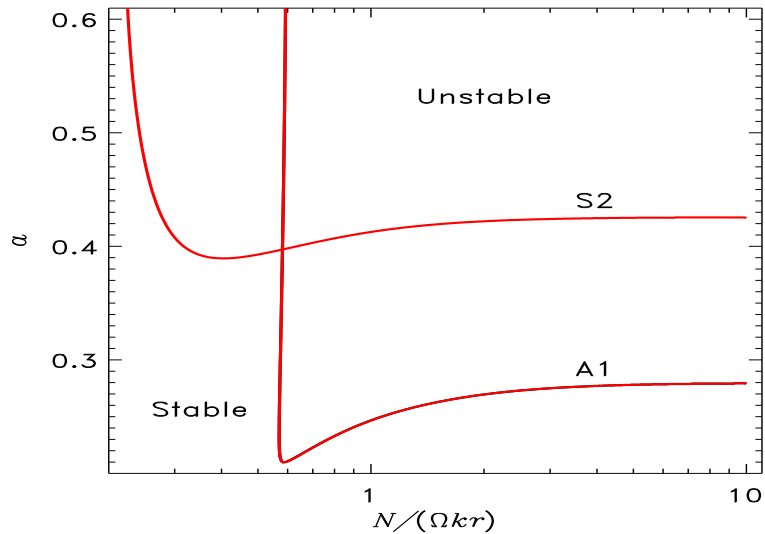}
   \caption{Stability map for  latitudinal differential
            rotation (3D hydrodynamics).  Most unstable are the 
	    perturbation modes  with the  vertical scale
	    $\hat\lambda \simeq 0.6$. The critical magnitude of latitudinal shear 
	    is reduced to 0.21 compared to the 0.29 value for  large wavelengths.  For  short wavelengths   the  instability for $m=1$ rapidly disappears by dissipation effects.
              }
   \label{Wf2}
\end{figure}

For positive $a$  in \eref{W28} the modes A1 and S2  become  unstable for sufficiently large $a=a_{\rm crit}$. Figure~\ref{Wf2} shows the dependence of the   $a_{\rm crit}$  on the normalized wavelength  $\hat\lambda$. Of course, for large enough radial wavelength  the   29\%-value of the Watson theory with 2D fluctuations  is reproduced. It is reduced, however,  to $a = 0.21$ in the full 3D model calculations. We see that in contrast to the original Watson approach short rather than long radial scales are preferred.  The minimum $a_{\rm crit}$ appears for   $\hat\lambda \simeq 0.6$,  so that the characteristic   wavelength of $\lambda \simeq 6$~Mm results for the solar tachocline. For shorter scales  the  instability for $m=1$ disappears due to the action of the dissipation processes but  modes with higher $m$ remain still unstable.  

\section{Concluding remarks}\label{discussion}
The presence of a meridional flow with an amplitude of  a few  meters per second at the bottom  of the convection zone is shown as necessary  for the magnetic tachocline  model. The flow provides the confined geometry (with field lines parallel to the outer spherical boundary) of the internal magnetic field which is necessary for the tachocline formation. The bottom flow is also a key ingredient of advection-dominated dynamo models for the solar cycle (see R\"udiger \& Hollerbach 2004 for detailed references). The flow was predicted theoretically  but its existence is not yet confirmed by observations. We  hope, of course, that helioseismology will probe  the deep meridional flow soon.

The polar cusps in the tachoclines of \fref{f9} are consequences of the assumed axisymmetry. This  assumption may not be realistic. The penetrating meridional flow is, however,  expected to confine even  nonaxisymmetric fields to the core so that the fields should   also be able to produce the  tachoclinic structures.  

The tachocline computations result in toroidal fields of 100 to 200 G, much stronger than the original poloidal fields. These toroidal fields can be subject to current-driven pinch-type  instabilities (Tayler 1973). Hence, the question is  whether the steady axisymmetric solutions of our model are stable  against all the  nonaxisymmetric disturbances. The threshold field strength for the instability were estimated to be of order $10^{2\dots 3}$~G (Spruit 1999; Arlt \etal 2007; Kitchatinov \& R\"udiger 2007) what is (only) somewhat larger than   the field amplitudes  in our model.

It is important for the magnetic tachocline model that the magnetic Reynolds number \eref{18} is large. This is true  for  microscopic diffusivity or for not too large eddy diffusivities up to about 10$^8$~cm$^2$s$^{-1}$. The tachocline should, therefore, be stable or only mildly turbulent to allow poloidal field confinement by meridional flow.
Some low level of turbulence may, however, be even necessary. The magnetic field is so efficient in producing   tachocline structures  that poloidal fields of only 1~G  lead to   radial scales  smaller than  1\% of  the solar radius. It is, however, $D_\mathrm{tach} \simeq 0.04R_\odot$. Hence, the internal poloidal field must   be (much) smaller than 1 G (Kitchatinov \etal 2001) or, if it is not, some kind of instability may prevent the tachocline thickness to reduce below the helioseismologically detected level. 

The stability issue is a clear perspective for future tachocline studies. As a corresponding application  we have developed in the last section    the 2D theory  of Watson (1981) for the hydrodynamic instability of latitudinal differential rotation   to a fully 3D theory. Now the critical wavelength of the unstable mode with $m=1$ is only 6 Mm while the critical  latitudinal shear (see \eref{W28}) is reduced from 29\% to 21\%.  This (linear) theory works with a Prandtl number of order $10^{-6}$ and a very high Reynolds number of order $10^{12}$.
\ack
The authors cordially acknowledge support by the Deutsche For\-shungs\-ge\-mein\-schaft
and by the Russian Foundation for Basic Research (project 05-02-04015).
\section*{References}
\begin{harvard}
\item[]
	Antia~H\,M, Basu~S and Chitre~S\,M
	1998 {\it Mon. Not. Roy. Astron. Soc.} {\bf 298} 543
\item[]
	Arlt~R, Sule~A and R\"udiger~G
	2007 {\it Astron. Astrophys.} {\bf 461} 295
\item[]
	Basu~S and Antia~H\,M
	1997 {\it Mon. Not. Roy. Astron. Soc.} {\bf 287} 189
\item[]
	Brun~A\,S and Zahn~J-P
	2006 {\it Astron. Astrophys.} {\bf 457} 665
\item[]
	Charbonneau~P, Dikpati~M and Gilman~P\,A
	1999a {\it Astrophys.~J.} {\bf 526} 523
\item[]
        Charbonneau~P, Christensen-Dalsgaard~J, Henning~R, Larsen~R\,M,
	Schou~J, Thompson~M\,J and Tomczyk~S
	1999b {\it Astrophys.~J.} {\bf 527} 445
\item[]
	Christensen-Dalsgaard~J, Gough~D\,O and Thompson~M\,J
	1991 {\it Astrophys.~J.} {\bf 287} 189
\item[]
	Dicke~R\,H
	1970 {\it Ann. Rev. Astron. Astrophys.} {\bf 8} 297
\item[]
	Ferraro~V\,C\,A
	1937 {\it Mon. Not. Roy. Astron. Soc.} {\bf  97} 458
\item[]
	Forg\'acs-Dajka~E and Petrovay~K
	2002 {\it Astron. Astrophys.} {\bf 389} 629
\item[]
	Garaud~P
	2007 Dynamics of the Solar Tachocline, Chapter 8 in
	{\it The Solar Tachocline.} Eds. D\,W~Hughes, R~Rosner and N\,O~Weiss
	(Cambridge: Cambridge Univ. Press) to appear
\item[]
	Gilman~P\,A
        2005 {\it Astron. Nachr.} {\bf 326} 208
\item[]
	Gilman~P\,A and Miesch~M\,S
	2004 {\it Astrophys.~J.} {\bf 611} 568
\item[]
        Kippenhahn~R and Weigert~A
	1994 {\it Stellar Structure and Evolution} (Berlin Heidelberg New York: Springer-Verlag)
\item[]
	Kitchatinov~L\,L and R\"udiger~G
	1999 {\it Astron. Astrophys.} {\bf 344} 911
\item[]
        Kitchatinov~L\,L, Jardine~M and Cameron~A\,C
	2001 {\it Astron. Astrophys.} {\bf 374} 250
\item[]
	Kitchatinov~L\,L and R\"udiger~G
	2006 {\it Astron. Astrophys.} {\bf 453} 329

\item[]
	Kitchatinov~L\,L and R\"udiger~G
	2007 {\it Astron. Astrophys.}  
\item[]
	Komm~R\,W, Howard~R\,F and Harvey~J\,W
        1993 {\it Sol.~Phys.}  {\bf 147} 207
\item[]
	Kosovichev~A\,G
	1996 {\it Astrophys.~J.} {\bf 469} L61
\item[]
	Kosovichev~A\,G \etal
	1997 {\it Sol.~Phys.}  {\bf 170} 43

\item[]
        Krause~F and R\"adler~K\,H
        1980 {\it Mean-Field Magnetohydrodynamics and Dynamo Theory}
	(Oxford: Pergamon Press) 

\item[]
	MacGregor~K\,B and Charbonneau~P
	1999 {\it Astrophys.~J.} {\bf 519} 911
\item[]
	Mestel~L
	1999 {\it Stellar Magnetism} (Oxford: Clarendon Press)
\item[]
        Miesch~M\,S, Elliott~J\,R, Toomre~J \etal
	2000 {\it Astrophys.~J.} {\bf 532} 593
\item[]
	Nandy~D and Choudhuri~A\,R
	2002 {\it Science} {\bf 296} 1671
\item[]
        Rempel~M 
        2005 {\it Astrophys.~J.} {\bf 622} 1320	
\item[]
        R\"udiger~G and Kitchatinov~L\,L
	1997 {\it Astron. Nachr.} {\bf 318} 273
\item[]
        R\"udiger~G and Hollerbach~R~L
	2004 {\it The Magnetic Universe} (Berlin: Wiley-VCH)
\item[]
	R\"udiger~G, Kitchatinov~L\,L and Arlt~R
	2005 {\it Astron. Astrophys.} {\bf 444} L53
\item[]
	Schou~J \etal
	1998 {\it Astrophys.~J.} {\bf 505} 390
\item[]
        Spruit~H\,C
	1999 {\it Astron. Astrophys.} {\bf 349} 189
\item[]
	Stix~M and Skaley~D
	1990 {\it Astron. Astrophys.} {\bf 232} 234
\item[]
	Tassoul~J-L
	2000 {\it Stellar Rotation} (Cambridge: Cambridge University Press)
\item[]
        Tayler~R\,J
	1973 {\it Mon. Not. Roy. Astron. Soc.} {\bf 161} 365
\item[]
        Watson~M
        1981 {\it Geophys. Astrophys. Fluid Dyn.} {\bf 16} 285
\item[]
        Weiss~N\,O
	1966 {\it Proc. Roy. Soc. London A} {\bf 293} 310
\item[]
	Wilson~P\,R, Burtonclay~D and Li~Y
	1997 {\it Astrophys.~J.} {\bf 489} 395
\item[]
	Zhao~J and Kosovichev~A\,G
	2004 {\it Astrophys.~J.} {\bf 603} 776
\end{harvard}

\end{document}